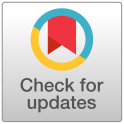

# Regulating human control over autonomous systems

Mikolaj Firlej
*Faculty of Law, University of Oxford, Oxford, UK*

Araz Taeihagh
*Lee Kuan Yew School of Public Policy, National University of Singapore, Singapore*

**Abstract**
In recent years, many sectors have experienced significant progress in automation, associated with the growing advances in artificial intelligence and machine learning. There are already automated robotic weapons, which are able to evaluate and engage with targets on their own, and there are already autonomous vehicles that do not need a human driver. It is argued that the use of increasingly autonomous systems (AS) should be guided by the policy of human control, according to which humans should execute a certain significant level of judgment over AS. While in the military sector there is a fear that AS could mean that humans lose control over life and death decisions, in the transportation domain, on the contrary, there is a strongly held view that autonomy could bring significant operational benefits by removing the need for a human driver. This article explores the notion of human control in the United States in the two domains of defense and transportation. The operationalization of emerging policies of human control results in the typology of direct and indirect human controls exercised over the use of AS. The typology helps to steer the debate away from the linguistic complexities of the term "autonomy." It identifies instead where human factors are undergoing important changes and ultimately informs about more detailed rules and standards formulation, which differ across domains, applications, and sectors.

## 1. Introduction

The development of unmanned vehicles and their increasing use in both commercial and military operations[1] has spotlighted the amount of effort that has been deployed toward greater autonomy in recent years. Yet despite significant progress in automation, the conceptual apparatus to address the effects of these developments remains rudimentary (Brundage *et al.* 2018).[2] In academic studies, the prevailing temptation has been to integrate the effects of new technologies into familiar doctrines, specifically by focusing on whether any particular innovation complies with the general principles of law.[3] Existing research on law and technology examines predominantly the questions of why actors follow particular rules and how to conceptualize varying degrees of compliance.[4] Little attention has been given to the assessment of how general policies are operationalized in practice by either administrative rules or practices.[5] Such studies may not only problematize the content of a policy but also provide insights into what areas require more detailed attention from rule-makers and standardization agencies.[6]

This article focuses specifically on the notion of human control, a public policy,[7] according to which humans should execute a certain significant level of judgment or control over highly automated or autonomous weapons (AWs).[8] Arguably the first official reference to "human control" was introduced relatively recently (2012) by the US government in the policy on AWs (Directive 3000.09; DoD 2012).[9] The US Department of Defense (DoD) directs that "autonomous weapon systems shall be designed to allow commanders and operators to exercise appropriate levels of human judgment over the use of force."[10] Despite the growing interest in the role of human judgment over the use of autonomous systems (AS) in various industries, the US Department of Transportation (DoT) has recently issued a key document (AV 3.0) which no longer assumes human control as necessary to AS, and thus supports the path towards full automation.[11] While AV 3.0 is not a formal policy of DoT, the document provides a framework for the integration of Autonomous Vehicles (AVs) into the US transportation system and essentially serves as *de facto* US policy on AVs. However, many states still argue that a human must retain control over AVs and such a policy should guide a work of developers and manufacturers.[12] Selected states have introduced legislation with an explicit reference to the notion of human control, for instance, in the form of safety drivers.[13]

This article explores the notion of human control over AS in the United States in the domain of defense and transportation. It explores emerging policy and regulatory approaches that offer more insight into the content of this vague notion. It presents a conceptual typology of human control wherein two main types are distinguished: (i) direct control, which is a level of specification on how dependent AS are to be on the human, for example, whether a human operator can override machine decision; and (ii) indirect control, which is a level of trust one can have in the performance of AS. This framework helps to pinpoint where human factors are undergoing important changes and informs about more nuanced rules and standards formulation, which differ across domains, applications, and sectors.

The remainder of this article is as follows: Section 2 deals primarily with methodological considerations. It situates our research focus on the operationalization of human control in the wider problem of "autonomy" of systems. Further, it presents the top-down model of policy analysis, the notion of operationalization, and why a comparative study offers relevant insights into our research question. It justifies our assessment of general policy first, before exploring more specific rules applicable to both the transportation and defense industries. Sections 3 and 4 apply the methodological framework to the programs and rules of the US government with respect to AWs and AVs, respectively. These two sections explore how the requirement of human control is defined by US government policy and examine various administrative measures that have been adopted by the military and by various states' administrations in the case of AWs and AVs, respectively, that demonstrate the content of this notion. Section 5 compares how the notion of human control is defined in the two domains and argues that the focus on operationalization of human control may better inform emerging governance approaches toward ASs, before the article concludes in Section 6.

## 2. Methodology

This section highlights the research methodology of this study. Section 2.1 presents the objective of our research by focusing on human factors, rather than the notion of autonomy and Section 2.2 explains how we conducted the operationalization studies through the comparative lens.

### 2.1. Research objective: focus on human factors not autonomy

This article focuses on the operationalization of human control in the domain of defense and transportation in order to provide insights into the typology of human control as such. We argue that the typology of human control helps to provide more clarity in the current debate on the future of human agency as the use of AS increases. It can also inform the work on the governance of AS by shifting the focus from the regulation of the vague notion of autonomy toward more detailed rules and standards applicable to the changing role of human factors.

The current debate on the future of AWs and AVs is often hindered by a lack of consensus over the definition of basic terms, particularly the definition of AS. The scope and characteristics of autonomy are uncertain. It is a challenge to agree on a definition of "autonomy" even when the concept is applied to humans.[14] These problems are further exacerbated when applied to machines, primarily due to uncertainty as to whether the concept of personal autonomy can be extended to other entities, and, if not, what is qualitatively different in the notion of personal autonomy that renders it incommensurable or incomparable with that of machines.[15]

In the literature, AS are often differentiated from automated and automatic ones.[16] The term "automatic" is usually associated with simple, mechanical responses to environmental input. The term "automated" refers to more complex, rule-based systems, while the term "autonomous" is reserved for machines that execute some kind of self-learning abilities.[17] However, in practice, many machines fall somewhere within a spectrum of autonomy.[18] One way of evaluating a machine's level of autonomy is by measuring its level of dependence on humans while executing the Observe, Orient, Decide, and Act (OODA) loop.[19] The greater the machine's ability to observe, orient, decide, and act on its own, the greater its autonomy. However, three factors influence the degree to which a machine is considered to be automatic, automated, or autonomous: (i) the frequency of operator interaction that the machine requires to function; (ii) the machine's ability to execute tasks despite environmental uncertainty; and (iii) the machine's level of responsiveness regarding various operational decisions required for the machine to complete its mission.[20] Practical examples of the spectrum of autonomy include the Society of Automotive Engineers (SAE) Levels of Autonomy in AVs[21] or Sheridan's 10-level rankings.[22] Both illustrate the absence of a clear division between automatic, automated, and AS.[23]

Moreover, a machine's autonomy can vary at each stage of the OODA, thus a more comprehensive ranking should take this fact into account.[24] Some machines represent a combination of various types of functions, each with a different level of autonomy. In the context of weapon systems, there are at least four functions to a weapons system: trigger, targeting, navigation, and mobility.[25] Some weapons, such as precision-guided munitions, have autonomous mobility, triggering, and in some cases navigation, while the targeting is not autonomous. Weapons such as the Samsung SGR-1 Sentry Gun, on the other hand, have an autonomous targeting function through motion detectors, although their navigation, mobility, and triggering functions are not autonomous.[26]

In short, there is little agreement over how to define AWs or AVs through a vague notion of "autonomy." The confusion pervades discussions about when an advanced technological system is "autonomous," and what the implications of autonomy in the use of such systems are. The technology is developing rapidly while the intellectual apparatus to analyze the effects of these innovations is still at the nascent stage, particularly from a regulatory perspective.[27] The current state hinders further intellectual progress to move beyond definitional terms to discussion about where specifically a potential regulatory intervention is needed, that is, whether there should be more regulatory emphasis at the level of system design, or at the level of operation by humans.

This article aims to steer the debate away from the linguistic complexities of the term "autonomy" and focuses instead on pinpointing where the human factor is undergoing important changes and therefore may require particular attention from the relevant stakeholders. Furthermore, the presentation of human control in the form of a typology allows the mapping of a plethora of already existing concepts into a single format.

### 2.2. The exploration of human control through the lenses of operationalization studies

The article focuses on defense and transportation because they represent, at least based on official statements, contrasting approaches to the role of the human in the use of AS. While DoD is committed toward greater automation, they have explicitly included a reference on human control of AWs.[28] Recent guidelines from DoT, in contrast, do not assume the requirement of a human driver in the context of the growing automation of ground vehicles.[29]

These contrasting perspectives may indicate different regulatory measures applicable to the role of humans in the use of AS. Moreover, both AWs and AVs have recently received a fair amount of media attention. The current debate on AWs is largely framed by a fear that AWs represent or could represent a new, dangerous category of weapons, fundamentally distinct from other weapons, and therefore could represent new challenges for international and domestic legal regimes. Recently, such voices have been raised by the world's leading robotics and artificial intelligence (AI) pioneers who called on the UN to ban the development and use of "killer robots," as they referred to them. Moreover, the majority of countries have explicitly distanced themselves from the development and use of such weapons.[30] The current debate on AVs is dominated (though not univocally) by a hope that AVs can meaningfully improve people's well-being.[31] These rather radical positions usually ignore the wider changes in human–machine interactions related to the use of AS. In this article, we flesh out how the role of human factors is described and what might be the applicable regulatory measures.

The article first explores how the notion of human control is defined in the US policies applicable to AWs and AVs, and then focuses on the analysis of various regulatory measures that have been adopted to specify the content of this notion. The assessment follows a top-down model of analysis that begins with the study of policy objectives and then looks at more detailed rules and procedures.[32] Such top-down models of policy analysis are often criticized in implementation studies, as they emphasize the ability of decision makers to produce unequivocal policy objectives, and to control the operationalization and implementation stages.[33] Critics suggest alternative models of analysis, such as (i) bottom-up approaches that see lower-level bureaucrats as the main actors in the creation of policies, or (ii) hybrid approaches that try to overcome the divide between the two.[34] This article, however, employs a top-down approach because it studies the process of operationalization of a policy, rather than its implementation.[35] The study of operationalization of policies is different from the study of policy implementation.[36] These terms are often confused in the literature. Policy implementation is the translation of public policy into *practical outputs*. Implementation focuses on the delivery of a wide array of services for the parties impacted by the particular policy. Service delivery usually is defined as any contact with the public administration during which customers seek or provide data, handle their affairs, or fulfill their duties. Usually, however, public policies, enacted or not as law, are too general or abstract to enable an effective service delivery. This is where the operationalization comes to play. Policies need to be made operational to enable service delivery. This is why policies often require the development of formal, more detailed rules by bureaucratic institutions. "Organisations and agencies are created to translate laws into operational rules," as argued by Dye.[37] The operationalization therefore is not a synonymous of a policy implementation, and this carries certain methodological implications regarding empirical material. Policy implementation analysis requires the exploration of both formal rules and social practices that lead to the delivery of a policy. These practices are referred to as "outputs," that is, the actions that the government actually performs. The study of policy operationalization is not, however, concerned with policy outcomes ("Outcomes" refers to the results that are caused by those outputs), whether for instance one particular policy is more effective than another. Rather, it is concerned with the exploration of how policy objectives are defined and construed by bureaucratic institutions, how administrators understand particular terms, where they put emphasis on, and what matters to them.

The objective of this study is precisely to investigate various measures employed by both DoD and DoT in order to extract more insights about the concept of human control. Whether operationalization studies should involve the service delivery and impact assessment of a policy depends on research objectives. Our goal is to investigate various measures employed by the US government in order to extract more general insight into the notion of human control as such. Thus, empirical accounts of social practices are not decisive. Rather, at this stage—where we present a typology of various dimensions of human control, not the effectiveness of one or another measure—we are primarily concerned with how government understands this notion of human control. Thus, a top-down approach is preferable for the operationalization analysis, as it starts with the assumption that policy objectives are set by central policymakers and specified in more depth by lower administration.

The main difficulty in the study of operationalization lies in delineating the boundaries of policy analysis and the boundaries of specification of a given policy.[38] Public policies are principled guides formulated by the government to achieve intended outcomes. The process of operationalization is about translating general policies into more specific rules specifying what needs to be done and how in order to achieve intended outcomes. Thus, the operationalization of a policy usually includes the combination of legislative and executive action, but predominantly it is the domain of regulatory law, that is, law promulgated by the federal, state, and local administrative agencies.[39] Public policies, on the contrary, are not particularly detailed for two main reasons: they are intended to be flexible enough to apply to various changing market and political circumstances; and some matters are too technical for policy-makers, so they are necessarily delegated to more specialized branches of administration.

Therefore, the exploration of a policy refers to the main public documents which outline the US government's broad objectives regarding human control in defense (Directive 3000.09) and transportation (AV 3.0). The analysis of these documents is supplemented by other relevant high-level plans that relate to US objectives regarding AWs and AVs and that were presented respectively by DoD and DoT.[40] The exploration of specific rules that operationalize the notion of human control in defense is based on rules adopted by the US Army with respect to the development of new weapons, as well as targeting and engagement procedures. Specifically, we refer to the Enclosures to the Directive on AWs,[41] US targeting rules in the Army Techniques Publication,[42] the DoD Joint Capabilities Integration and Development System (JCIDS) documentation,[43] selected publications of the Defense Science Board,[44] and US weapons review rules. In the transportation domain, we present an analysis of selected state legislation, as transportation specific rules are predominately set by state, rather than federal legislature.[45] Our analysis is based on publicly available documents obtained from 2012 and 2018, from the adoption of the Directive on AWs and the first state legislative efforts to regulate AVs (both in 2012) to the publication of the DoD National Defense Strategy and DoT AV 3.0.

The limitations of this study are primarily related to the limitations of our sources. We relied only on publicly available documents. In particular, some military rules, such as the Rules of Engagement (ROE), are classified and may contain additional information regarding the specific operational tasks of human operators.[46] Moreover, further studies on empirical accounts of how various agents respond to operational rules could provide additional insights into the typology of human control over AS and further inform the debate about the governance of AS.

## 3. Human control in AWs

This section presents a case study of US regulatory measures applicable to the notion of human control over AWs. Section 3.1 explores how the notion of human control is defined by the US government, while Section 3.2 focuses on the analysis of various measures that have been adopted by the US administration, which demonstrate the content of this notion.

### 3.1. The notion of human control in AWs

Robotic systems with greater autonomy have been highlighted as a key component of the US military by all major documents outlining US defense and security priorities.[47] In the United States, the research, development, and deployment of AWs are specified by the DoD Directive 3000.09, *Autonomy in Weapon Systems* (2012). The DoD directive establishes national policy on AWs.[48] The directive specifies that AWs are those which, upon automation, can select and engage targets without human intervention.[49] The directive allows the development and full use of AWs to apply nonlethal, nonkinetic forces, such as electronic attack against material targets. However, in special cases, AWs can be used more broadly, and the directive does not apply to all parts of the US government or to cyberspace systems for cyberspace operations.[50]

The directive states that AWs and semi-AWs should be designed to allow commanders and operators to exercise "*appropriate* levels of human judgment over the use of force."[51] Interestingly, the United States does not explicitly refer here to the norm of human *control*, as they argue that the notion of control can be misleading and too restrictive. It may imply a form of *direct control* in the sense that AWs would be mere tools dependent on constant human input.[52] The directive, on the contrary, concentrates primarily on the stage of weapon's design. The directive states that AWs will need to go through rigorous hardware and software verification and validation (V&V) and realistic system developmental and operational test and evaluation (T&E), while training, doctrine, and tactics, techniques, and procedures (TTPs) are being established. These measures, according to policymakers, will ensure that AWs will function properly. The division between human judgment at the level of design versus direct control has also been emphasized by a US delegation to the Group of Governmental Experts on AWs:

(…) an operator might be able to exercise meaningful control over every aspect of a weapon system, but if the operator is only reflexively pressing a button to approve strikes recommended by the weapon system, the operator would be exercising little, if any, judgment over the use of force. On the other hand, judgment can be implemented through the use of automation.[53]

However, the directive in fact expands the notion of human judgment beyond the design stage in a situation where a weapon is unable to complete engagements consistently with operator intentions. In such cases, a weapon should seek additional human operator input before continuing the engagement or terminate engagements. We have here a classical reference to direct control.

Thus, one might argue that two dimensions of human control emerge. The first type is control-by-design. While the directive provides various procedures to ensure "the appropriate level of human judgment," it does not specify what this "appropriate level" constitutes. Next section will explore this issue in the light of operational rules adopted by the US military administration. The second type of control is related to the ability of a human to directly exercise control by terminating an AW's engagement. This "finger on the button" control presents a different dimension of control, even though prior design may determine it. It is different because a designed mechanics of control is determined *ex-ante* engagement, while a "finger on the button" is the ability to exercise control *ex-post* engagement, that is, in a realm of warfare during a real-time operation.

This difference has two main consequences for operational rules. First, regulation of human control-by-design is primarily concerned with the notion of trust, that is, whether actions by an artifact or designed object can be deemed trustworthy. Trust is critical for a fighter in a hostile environment. As the Defense Science Board (DSB) found, trust in AS is "core to the DoD's success in broader adoption of autonomy."[54] Trust in the context of AWs is often defined by the reliability and predictability of a system.[55] It is characterized by a pattern of behavior observed by the trustor and the belief that the recipient of trust (the AS) will respond in a particular way.[56] As pointed out by Roff, this type of trust persists in the absence of any detailed knowledge or understanding of the "inner workings" of the machine's design, structure, or past history.[57] An operational consequence in the context of AWs is that there must be certain rules that ensure a significant level of reliability and predictability in such weapons in relation to human interface. Second, regulation of human control in the form of "finger on the button" is primarily concerned with an emergency of last resort. It is an insurance policy, in case something goes wrong. In academic literature, this type of human control is known as "human-on-the-loop" and refers to the ability of a human to override a machine's action.[58] It differs from "human-in-the-loop," which refers to those machines that can select targets and deliver force only with a human command, and also from "human-out-the-loop," that is, robots that are capable of selecting targets and delivering force without any human input or interaction.

While the distinction is clear, difficulties arise in practice. Consider the following scenario. Weapon A can act autonomously. A human has been involved initially in framing the selection and engagement with target, but ultimately the decision on a specific target is delegated to a weapon. After the initial selection, a human cannot override a machine's decision. This is an example of "human-out-the-loop." Weapon B can also act autonomously.

Similarly, a human has been involved initially in framing the selection and engagement with the target but ultimately the decision on a specific target is delegated to a weapon. After the initial selection, the human does have the ability to override a machine's decision; however, she has a very restricted time in which to veto engagement. Should Weapon B be considered to be controlled by a human? A challenge here is to present the conditions in which a human is *actually* able to exercise her right (or duty) to override a weapon's decision to engage. This is not just a technical question. Consider, for instance, a scenario in which Weapon B is used in operation against an adversary with an AW. In a dynamic and unstructured environment, Weapon B was not accurately responsive and engaged beyond the initial selection of targets. However, an adversary does not have any limitation when it comes to the termination of engagement other than the final destruction of the opponent. The question is what legal, ethical, and technical criteria should a human operator consider in such situations when exercising her right (or duty) to override a machine engagement? Hence, an operational consequence in the context of AWs is that there must be certain guidelines as to the circumstances in which a human can override a weapon's engagement.

### 3.2. Operationalization of human control over AWs

This section focuses on the analysis of various measures that have been adopted by the US military administration that demonstrate the content of the notion of human control. As discussed in the previous section, there are two formulations of such notion:

1. Human control-by-design, formulated in the directive as: "The system design incorporates the necessary capabilities to allow commanders and operators to exercise appropriate levels of human judgment in the use of force."[59]
2. Human control as a "finger on the button," formulated in the directive as: "The system is designed to complete engagements in a timeframe consistent with commander and operator intentions and, if unable to do so, to terminate engagements or seek additional human operator input before continuing the engagement."[60]

The operational consequences of human control-by-design focus on ensuring a significant level of reliability and predictability in such weapons in relation to human interface. The directive stipulates that before fielding, each weapon should be a subject of legal review and – in addition – AWs will need to go through rigorous hardware and software V&V and realistic system developmental and operational T&E, while training, doctrine, and TTPs will have to be established. Therefore, two types of general procedures shall be used to gain the necessary trust at the level of design: (i) legal review and (ii) technical evaluation procedures.

The weapons review was established in Article 36 of the 1977 Additional Protocol 1 to the Geneva Conventions [Protocol 1]. While the United States is not officially a party to Protocol 1, they have established a national weapons review, and they consider the review obligation to be customary in international law.[61] Although overall responsibility for conducting legal reviews lies with the DoD, all of the military services have a department responsible for conducting legal reviews. In order to ensure alignment across US military services, military lawyers work with the DoD's Law of War Working Group. The most important element of weapons review is a compliance test with laws of armed conflict. It is widely accepted that the assessments must consider: (i) Is the weapon, means, or method of warfare prohibited by a specific treaty or customary law provision? (ii) Is the weapon, means, or method of warfare inherently of a nature to cause superfluous injury or unnecessary suffering in its regular or intended use? (iii) Is the weapon, means, or method of warfare inherently of a nature to be indiscriminate, in its normal or intended use?[62] Depending on the response to these questions, the authority conducting the review could place restrictions, approve with conditions or make recommendations on the use of new weapons, which could be integrated into the ROE and practical training. In the case that a new weapon does not fulfill all requirements posed by weapons review, the reviewer may state corrective actions identified during the process that should be followed in order to ensure that a subsequent review process would allow the envisioned weapon to pass the legal review. During this stage, the reviewer considers whether legal restrictions on the use of the weapon apply or whether practical training or ROE measures specific to a particular weapon system are necessary. In academic literature, there is no agreement on whether currently existing AWs meet these requirements, but many authors argue that existing AWs are not necessarily unlawful *per se* as they do not automatically violate any principle of international armed conflict.[63] However, it is very likely that such weapons might be used in an unlawful manner, depending on circumstances. It is particularly unlikely, especially given current technological developments, that AWs will be able to conduct sensitive, context-dependent, and standard-based assessments of proportionality and military necessity.[64] The above issue, however, could not be addressed without a thorough examination of the empirical evidence and specific technical considerations.[65] Therefore, a weapon system's V&V, T&E, and TTPs will inform the substantive legal review significantly.

In 2011, the United States published a report in which it stated that growing automated functionality would "introduce unexpected levels of risk" and challenge current T&E and V&V capabilities.[66] A second roadmap (2013) discusses testing in much greater detail and articulates standards of testing.[67] In this document, 'trust' refers explicitly to the satisfaction that an AS will operate in an expected or predictable manner. Current T&E and V&V capabilities, effectively institutionalized in the DoD JCIDS, are based on the safety assurance concept. According to this concept, the primary objective of the JCIDS process is to ensure that the capabilities required by joint warfare[68] are identified, along with their associated operational performance criteria, in order to successfully execute the missions assigned. It means that before any new capability can enter the development process related to reviewing and validating its requirements, the originating sponsor organization (a weapon manufacturer) is obliged first to identify the AWs system capability requirements related to its functions, roles, mission integration, and operations. It must then determine if the pursuit of such capability presents an unacceptable level of risk relative to potential operational benefits.[69] Obligatory assessments and reviews are captured in the Capability Development Document, which addresses system safety in accordance with current DoD guidance. However, AWs present a number of challenges with regard to existing T&E practices.[70] Currently, the majority of systems are continuously monitored by humans who can detect deviations from desired performance and correct the behavior of the system. AI, on the contrary, enables increasing *cognitization* of machines, which not only makes them faster or smarter than humans, but increases the unpredictability of their actions due to self-learning and emergent coordination capabilities. Ensuring that such a system will respond appropriately to all possible inputs will exceed the capability of conventional testing.[71] Self-learning systems exhibit different behaviors, as they incorporate data about their tasks from a variety of sources in real time and learn to provide better results partly based on *their own experience*. Therefore, such software cannot be exhaustively tested. While the commercial sector offers certain guidelines when it comes to evaluating and qualifying such systems, a lack of specific standards for continuous V&V for AWs does not provide sufficient reassurance when it comes to the reliability and predictability of such weapons in relation to human interface. This statement is supported by the *Unmanned Systems Integrated Roadmap FY2013–2038*, which finds that "existing ranges and test facilities have been adequate to test systems with very limited autonomous capabilities."[72] Similarly, the *Summer Study on Autonomy* concludes that current conventional testing capabilities are "inadequate for testing software that learns and adapts."[73]

Moreover, Section 2 of Enclosure 3 of the directive states that the undersecretaries of defense for policy and acquisition, technology, and logistics may request a Deputy Secretary of Defense waiver for T&E and V&V requirements, with the exception of the requirement for a legal review, in cases of urgent military operational need.[74] It means that only weapon review is effectively a binding requirement that informs on operational trust at the level of design.

The operational consequence of human control as a "finger on the button" is that there must exist certain guidelines under which a human can override a weapon's engagement. As the directive states, a human operator can override a weapon when the system does not complete engagements in a timeframe consistent with operator intentions. However, there are no specific guidelines with respect to how a human should execute this function. One may argue that the human operator is a highly experienced agent, so she would be able to understand the system's limitations. The empirical research on human-machine interface suggests, however, that humans often over- or under-estimate a machine's capabilities.[75]

Moreover, humans suffer from "automation bias," where they accept a machine's decision even though it is not correct.[76] Automated or autonomous decision-making may have also detrimental effects on the human ability to perform the same tasks as the machines. The increasing delegation of tasks to machines may lead to a degradation in skill levels, which might become a problem, particularly in safety critical applications.[77] Therefore, the increasing reliance on autonomous capabilities will require changes in existing command-and-control concepts in order to ensure a greater level of trust. Irrespective of these deficiencies, Section 2 of Enclosure 3 of the directive also provides an opportunity to waive the requirement of human control as a "finger on the button" under the same circumstances as a waiver for T&E and V&V requirements, that is, in cases of "urgent military operational need."[78] While it is understandable that all of DoD's needs cannot be met by the same acquisition processes, there is a clear recommendation from DSB that "rapid acquisition processes" must be based on proven technology and robust manufacturing processes. Otherwise, attempting to squeeze new technology developments into an urgent timeframe creates significant risks, particularly when it comes to the operation of self-learning systems. The directive does not define "urgent military operational need," which makes the phrase vulnerable to broad interpretation, whereas DSB found that current approaches to implementing rapid responses to urgent needs have not been sustainable.[79]

Many academics argue today that AWs should be guided by human control or that a human should somehow retain control over AWs. Our research shows that US policy indeed includes a requirement for human control over AWs as a guiding principle. The operationalization of human control reveals its two dimensions: human control-by-design and human control as a "finger on the button." It is, however, unclear under what conditions a human can override a weapon when the system does not complete engagements in a timeframe consistent with operator intentions. Both types of human control are required during the operation of such weapons, but in the context of urgent military need AWs may only satisfy the requirement for control-by-design. While there are many specific requirements for weapons design, effectively, only weapon review is a binding requirement that informs on operational trust at the level of design. Table 1 provides a summary of the operationalization of human control over AWs.

## 4. Human control in AVs

This section presents a case study of US regulatory measures applicable to the notion of human control over AVs. Section 4.1 explores how the notion of human control is defined in the US policy, while Section 4.2 focuses on the analysis of measures that have been adopted by selected states, which demonstrate the content of this notion.

### 4.1. The notion of human control in AVs

The US government has not established an explicit federal policy on AVs.[80] It has been the deliberate choice of federal policy-makers to issue only voluntary guidance to states called *Preparing for the Future of Transportation:*

**Table 1**　Human control over AWs

|  | Type of control | Engagement role | Intervention role | Key requirements |
|---|---|---|---|---|
| Human control as "Finger on the Button" | Human in the loop | Human | Human | (i)　The active presence of a human operator, (ii)　system's responsiveness to complete tasks in a timeframe consistent with operator and commander intentions |
|  | Human on the loop | System | Human | (i)　The presence of a human operator; (ii)　a manual override feature that allows an operator to assume control of the AS at any time |
| Human control by design | Design control as default | System | System or human | (i)　Relevant software V&V, T&E, and TTPs, (ii)　compliance with relevant law, BUT in the case of "urgent military need" only weapon review applies |

*Automated Vehicles 3.0* (AV 3.0).[81] AV 3.0 provides a framework for the safe integration of AVs into the US surface transportation system and supplements the previous *Automated Driving Systems: A Vision for Safety 2.0* (ADS 2.0). The DoT does not generally rule out support for the full automation of all aspects of the dynamic driving task as long as AVs are developed according to performance-oriented, consensus-based, and voluntary safety standards.[82] A key objective of AV 3.0 is to relax assumptions in existing regulations related to the presence of a human.

Given that the AS rather than the human driver carries out the dynamic driving tasks for an AV, it is unclear which entity may be considered the "driver" or to be "controlling" the vehicle.[83] The guidelines seek to address these ambiguities, stating that definitions of "drive" and "operator" will be adapted to allow for these entities to refer to both a human as well as an automated system.[84] By doing so, federal vehicle regulations will "no longer assume" that a driver is human or that a human would always be present onboard a vehicle during its operations.[85] Specifically, the DoT suggests new rulemaking for AVs, which would include more flexible standards for AVs.[86]

The DoT expressed their commitment to promote the adoption of automated driving system (ADS)-equipped vehicles that represent Levels 3–5 of vehicle automation, as defined by the SAE.[87] The majority of ADS-equipped vehicles are at a Level 2, which requires a human driver to keep their hands on the steering wheel. Some car manufacturing companies are working toward Level 3 vehicles that allow drivers to take their hands off the steering wheel and feet off the pedals but return control to the driver when the car requests it. Other manufacturers have considered moving straight from Level 2 to Level 4, where an ADS performs all aspects of the DDT even if a human driver does not respond. Level 5 represents full automation. In the remaining part of this article, we refer to AVs as vehicles at Levels 3–5 as defined by SAE. This is consistent with the AV 3.0 guidance that refers to AVs separately from human-driven vehicles.[88]

While AVs are advancing rapidly, the technology is still not yet fully reliable, due to existing limitations that could increase the risk of collision if a human driver is unable to regain control in a timely manner. Fatal accidents have already occurred.[89] While most autopilots are still at Level 2 and are considered as a driver-assistance tool, not a driver replacement, their technological limitations are significantly affecting performance while creating an illusion of confidence for many drivers.[90] Effective handover of control to the human driver is thus essential to mitigate the safety risks resulting from unpredictable technological failures, at least during the transition from manual to fully automated AVs.[91] However, the conditions under which humans should regain control over AVs and *vice versa*, as well as exactly how the AV system should "prompt" its human passengers to take over control, are still unclear.[92] The design of the human–machine interface is critical as it shapes the ways in which the AV system prompts the human driver to regain control and thereby, the human's performance in responding during handover scenarios.[93] Further, the reduced level of human control over the decision-making of the vehicle's raises questions regarding whether and how AVs should be programmed in advance to avoid causing harm or damage, particularly during unavoidable accidents.[94]

The US federal guidelines published by the DoT seek to address these risks, considering various dimensions of human engagement in the context of AVs. The first notion of human control is "hands on the wheel," which relates to the ability of a human to take over the DDT. This form of control differs from a situation where a human driver is responsible for the entire DDT, in other words when a human is actively driving a car. Several existing National Highway Traffic Safety Administration (NHTSA) safety standards for motor vehicles assume that a human occupant will be able to control the vehicle, and many standards incorporate performance requirements and test procedures geared towards ensuring safe operation by a human driver.[95] However, both DoT and the NHTSA consider the possibility of setting exceptions to certain standards (relevant only when human drivers are present) for ADS-equipped vehicles.[96] In the next section, we explore how selected states consider this notion of human control, by examining whether they require a human driver to be present, and if so, under what conditions.

The second notion of human control over AVs is a form of control-by-design, which mirrors the case of AWs. In the context of AVs, two sub-categories of this concept emerged. Firstly, control-by-design relates to the reliability and predictability of a system's design. AV 3.0 guidelines recommend a number of measures that concentrate on hardware and software V&V to build confidence in the use of AVs and mitigate safety risks. AV 3.0 specifies the following conditions:

U.S. DoT envisions that entities testing and eventually deploying ADS technologies will employ a mixture of industry best practices, consensus standards, and voluntary guidance to manage safety risks along the different stages of technology development.[97]

Interestingly, instead of specifying any requirements for AVs, the DoT aims to promote discussion around various visions for ensuring safety and managing AVs safety risks, and the AV 3.0 guidelines refer to a variety of measures at different stages of development. However, unlike the design requirement of AWs discussed earlier, AV 3.0 is not oriented towards an ultimate goal whereby the system design should incorporate the necessary capabilities to allow the human operator or driver to exercise appropriate levels of human judgment in the use of an AV. While safety drivers are still expected to be in the loop, particularly to ensure safety during road testing, and identify new scenarios of interest, they mainly serve as a secondary risk mitigation strategy and will not be primarily relied upon to ensure safety in the future, when AVs will mainly operate without human drivers.

The second subcategory of control-by-design relates to the ethical principles that AS follow in situations where ethical judgment is required, such as unavoidable accidents. The AV 3.0 guidelines recommend additional software assessments during situations of handling failure, crash-imminent scenarios, and edge cases, but do not specify how such systems should behave during unavoidable accidents. While this debate is still in its nascent stages, developers will have to decide which ethical principles should be used to inform the AV's decision-making during such incidents. For instance, whether the AV should adopt more a consequentialist or a deontological approach remains an open question.

In the next section, we explore the operational consequences of both notions of human control – control in the form of "hands on the wheel" and control-by-design – in the context of AVs. Under federal law, no state or local government may enforce a law on the safety performance of a motor vehicle or motor vehicle equipment that differs in any way from the federal standard. However, in the current situation – in the absence of law – states have adopted various measures to ensure the safe use of AVs. Thus, the next section does not refer specifically to one state, but rather compares the various approaches of selected state legislatures.

### 4.2. Operationalization of human control over AVs

This section focuses on the analysis of various measures that have been adopted by states, which reveal the content of a human control notion. As discussed in the previous section, there are two formulations of this requirement:

1. Human control in the form of "hands on the wheel" as the manual ability to intervene and take control of DDT. This requirement is formulated in AV 3.0 primarily through the requirement for a "safety driver" during the testing phase.
2. Human control-by-design that considers the predictability and reliability of a system as well as the ethical principles that shape the AV's response during situations that require ethical judgments. AV 3.0 recommends testing and evaluating AVs by adopting a mixture of industry best practices, consensus standards, and voluntary guidance, but it does not specifically refer to any ethical principles.

The notion of human control as the manual ability to intervene is operationalized through the requirements with respect to the presence of human driver and the conditions under which a human driver is required. Based on the review of state legislations, there are currently four main approaches:

a  States that are silent as to whether a human driver is required. Examples include New Jersey, New Hampshire, and Montana, among others.[98] The majority of these states, however, actively discuss the adoption of AVs in the near future.
b  States, such as Illinois,[99] Connecticut,[100] and Washington D.C.,[101] that formally require a human driver to be prepared to take control of the AV. AVs may operate on a public roadway provided that the vehicles have a manual override feature allowing a driver to assume control at any time and have a driver seated in the control seat of the vehicle who is prepared to take control of the AV during operation. Human drivers are required to have a valid driver's license. However, none of these states specify how many and which of the vehicle's tasks must actually be undertaken by the driver and under what conditions.
c  States that do not require the presence of a human driver in the vehicle but require remote human control of the AV. California State has recently passed a law according to which car manufacturing companies are required to link their cars to remote operators. They are responsible for monitoring multiple cars and taking over their controls, if needed.[102] Specifically, they must establish continuous monitoring of the vehicle and two-way communication links, descriptions of how the manufacturer will monitor the communication link, and explanations of how all AVs will be monitored. The manufacturers are allowed to hire third-party companies to handle remote operation, but some of them have already developed their own remote-driving technologies.[103]
d  States that do not require the presence of a human driver in the vehicle. Florida has recently amended their regulations to allow operation of AVs on public roads and to eliminate requirements for a human driver in the vehicle.[104] Other states in this position include Michigan, Tennessee, Arizona, and Georgia.[105]

States that do not require the presence of a human driver stipulate various specific requirements to mitigate safety risks. Three main requirements are usually considered by state legislators:

a  Design-related requirements—there are usually two sub-categories of this requirement. First, the human driver need not necessarily be present in the vehicle, as long as the vehicle is capable of achieving a minimal risk condition, particularly in the context of malfunctioning.[106] California requires vehicles to have steering wheels and brake pedals. If automakers want to test cars without such controls, they will need to receive a waiver from the NHTSA.[107] Moreover, Tennessee and California require AVs to have "Automatic crash notification technology," a wireless communications and vehicle location technology to determine the need for emergency response in the event of a vehicle accident.[108] They also require a data-recording system that is capable of recording the ADS's status and other vehicle attributes, including speed, direction, and location, during a specified time period before an accident, as determined by the car manufacturer.[109] Furthermore, car manufacturers are required to meet applicable industry standards and information privacy laws (Lim and Taeihagh, 2018). Second, the vehicle should be capable of operating in compliance with the applicable traffic and motor vehicle safety laws and regulations of the state that govern the performance of the dynamic driving.[110]
b  A permit requirement – a majority of states, such as Florida, California, and Nevada, require car manufacturers to obtain a permit to operate AVs on public roads.[111] Currently, the only exception is Arizona.[112]
c  A driver's license requirement—some states such as Tennessee, Massachusetts,[113] and California require the AV's operator to possess a valid driver license, while states such as North Carolina specify that a driver's license is not required for an AV operator, but they require an adult to accompany a child under 12 years in the vehicle.[114]

While some states specify many design requirements, there is a lack of reference to the idea that design should incorporate the necessary capabilities to allow human operators to exercise appropriate levels of human judgment in the use of the cars, as such a design requirement is applied to AWs. Instead, various design-specific requirements are in place to mitigate safety risks, while advancements in driverless technology increasingly eliminate any kind of human input. As an example, in Arizona and California automakers are already able to manufacture and operate cars without a steering wheel, brake pedal, or accelerator.[115] These innovations are supported by the DoT, which is committed to the wider development of AVs by taking a soft approach to regulatory requirements. The DoT proposed self-driving rules that focus only on a set of guidelines calling for automakers and technology companies to voluntarily report on their testing and safety of AVs to the NHTSA before public deployment.

Moreover, they exempt AV developers from meeting many existing transport safety rules, including the requirement that a human operator should be inside the vehicle and that vehicles have safety features such as a steering wheel, brakes and mirror that are functional at all times. Whether the engineering process will be oriented towards ensuring a certain level of human oversight, similar to the case of remote control in California, is uncertain. The DoT, like individual states, does not provide any detailed requirements relating to software and hardware V&V processes. There are no requirements for a minimum "vision test" for the AV system to assess whether the AV is able to properly identify its surroundings, including other cars, pedestrians, and traffic signs, and respond appropriately. However, validation is critical to guarantee the safety and reliability of the system's performance. While there is an international safety standard for conventional cars (ISO-26262), its use is not currently mandated by the US government. If accepted, a framework such as ISO-26262 might develop into the standard engineering process requirement to validate, verify and ultimately approve the operations and safety of AVs. However, ISO-26262 only addresses functional safety, such as hardware and software developed to minimize risks caused by a functional failure but does not address design flaws in algorithms. This standard is referred to as *The Safety of The Intended Functionality* and is included in SO/PAS 21448. Currently, there are no requirements for transparency in the design of algorithms underlying the AV system, which is crucial to identify the causes of crashes or system malfunctions and the appropriate solutions to these problems. A problem widely known as "opacity," which characterizes many machine learning (ML) systems (Surden and Williams, 2016), is a major challenge in determining aspects of an AV's decision process that contributed to a crash.[116] Such data analysis practices exist in commercial aviation.[117] Whether safety is ensured within the scope of existing standards or is supported otherwise, it is essential for independent third-party assessors to review existing safety procedures to ensure that vehicles are as safe as they need to be for use on public roads,[118] which are also not yet required by states' existing regulations. In addition, there is a lack of AV-specific ethical guidelines available to AV developers for informing an AV's responses during inevitable crash situations.[119] Regulators' reluctance to specify more requirements for automakers can be attributed to concerns over stifling innovation at the current nascent stage of AV development and the lack of consensus around the methods that should be used to validate AI and ML systems.[120]

Many academics today argue that AVs should be guided by human control or that a human should retain some degree of control over AVs. Our research shows that while federal US policy considers various dimensions of human control over AVs (manual control, design control's default, and control by "ethical code"), it no longer assumes that such control is necessary as the full AS replaces the role of the human driver in conducting the DDTs. Operationalization of these dimensions of human control by states reveal that states are primarily interested in safety of AS, rather than retaining elements of human control. Human control in the form of the human driver's physical presence is seen primarily as a safety backup of a rather temporary nature. Moreover, control-by-design of the AV system does not require capabilities to allow human operators to exercise appropriate levels of human judgment over the vehicle's operation. Lastly, California was the first state to introduce a combination of manual and design-type control in the form of remote control. Remote control represents a manual type of control because it satisfies the "human-on-the-loop" condition, which refers to the ability of a human to override a machine's action. Remote control also imposes specific design requirements, such as two-way communication links and technical explanations on the ways in which the AV will be monitored. The governance model adopted in California is therefore similar to the one applied to AWs in the US: it captures the requirements of predictability and reliability of a system (design) with an insurance policy in the form of human control as a last resort (manual). Table 2 shows the summary of the operationalization of human control over AVs.

**Table 2**  Human control over AVs

|  | Type of control | Driving role | Intervention role | Key requirements |
|---|---|---|---|---|
| Human control as "Hands on the Wheel" | Human in the loop | Human | Human | (i) The active presence of a human driver with a driver's license |
|  | Human on the loop | System | Human | (i) The presence of a safety driver; (ii) vehicles have a manual override feature that allows a driver to assume control of the AVs at any time; (iii) a valid driver's license |
|  | Remote control | System/Human | Human | (i) AV companies are required to link their cars to remote operators; (ii) remote operators with a driver's license; (iii) Two-way communication links; and (iv) monitoring information |
| Human control-by-design | Design control as default | System | System or human | (i) Relevant software V & V and T & E; (ii) compliance with relevant law; (iii) automatic crash notification, technology; (iv) wireless communication; and vehicle location technology; (v) a data recording system; (vi) A permit to operate AVs; and (vii) a valid driver's license |
|  | Control by "ethical code" | System | System | (i) Same requirements as per design control as default; (ii) additional requirement in the form of ethical guidelines for IT developers |

## 5. Discussion: typology of human control and implications for the governance of AS

The operationalization of human control in the domain of defense and transportation provide insights into the typology of human control. The typology of human control helps to provide clarity in the current debate on the future of human agency as the use of AS increases. It steers the debate away from the linguistic complexities of the term "autonomy" and focuses instead on pinpointing where the human factor is undergoing important changes and therefore may require particular attention from the relevant stakeholders. Further, the presentation of human control in the form of a typology allows the mapping of a plethora of already existing concepts into a single format (Table 3). Below, based on the analysis in Sections 3 and 4, we provide a classification of human control over AS:

1 Direct control consists of two types:
   a Human-in-the-loop control is characterized by constant and uninterrupted physical control over the system. A human is always actively engaged with a system and consequently influences its outcome. In transportation, an example would be a human driver who drives a vehicle.
   b Human-on-the-loop control is characterized by the ability of a human to override or intervene in the action of a system. It is usually considered as a measure of last resort. In the context of transportation, we referred to this notion as control in the form of "hands on the wheel" in the military as a "finger on the button" type of control.

   A remote control is a specific type of direct control, which can be used as either a human-in-the-loop or human-on-the-loop type of control, depending on technical specification.[121] This type of control refers to the operation of a device from a distance, usually wirelessly. In transportation, such control is required for AVs in California in the form of human-on-the-loop control, while in the military, there are unmanned aerial vehicles which are remotely controlled and can represent either the human-in-the-loop or human-on-the-loop type of control.

2 Indirect control or control-by-design consists of two dimensions:
   a The necessary capabilities to allow human operators to exercise appropriate levels of human judgment in the use of the system. This requirement usually refers to the ability of a system to be both predictable and reliable, and specifically relates to the software and hardware V&V process.
   b The necessary ethical guidance that AS follow in situations of inevitable accidents. This dimension of control is particularly human because it involves deliberate choice as to the moral guidance of AS in a situation when crash or accident is inevitable. Currently, this dimension of control is relatively undeveloped, particularly in the light of existing regulations.

The operationalization of human control, which resulted in the above typology, can provide insights into the wider debate on the governance of AS. The current debate is largely focused on the formulation of general policies and principles that should guide the safe development and use of AI. One of such general policies is a widely discussed policy of human control. However, a more rigorous understanding of human control through the lenses of operationalization research helps to identify where human factors are undergoing important changes and ultimately inform about more nuanced rules and standards formulation, which, as the article shows, might differ across domains, applications, and sectors. Rules are the most constraining and rigid legal norms. Standard are less constraining than rules as they are consensus-driven guidelines that are approved by a recognized body aimed at the achievement of the optimum degree of order in a given context.[122] The standards are considered as a type of regulatory tool particularly suitable for addressing ever-changing realm of technology development.[123] Over the recent decades, they have gained increasing attention from governments as they typically allow better coordination of activities in both in domestic and global markets.[124] This has been recently confirmed by the White House "Executive Order on Maintaining American Leadership in Artificial Intelligence" (2019), which states that the United States should lead the work on developing international *technical standards applicable to AI*.[125] Currently, the process of standardization is conducted by the wide array of international organizations. Among them the most active in the domain of AI are The International Organization of Standardization (ISO) and The International Electrotechnical Commission (IEC), who work together in a joint committee (ISO/IEC JTC 1/SC 42) as well as The Institute of Electrical and Electronics Engineers Standards Association (IEEE).[126] While there is certain progress already underway, particularly with respect to establishing standards applicable to algorithmic bias and trustworthiness, there is still much work to be done particularly with regard to "foundational standards," that is key definitions, taxonomies, and areas of focus.[127] The main purpose of these efforts is to provide a common language for regulators and all parties interested in developing AS.[128] The work draws inspiration from the earlier deliberation on big data which resulted in establishing the relevant standard (SO/IEC 20546:2019) as well as use cases.[129] This standard provides a set of terms and definitions that serve as a foundation for all big data-related standards. In the domain of AI, there are notable challenges related to standardization of language and, as a result, one can observe that new AI-related standards continue to emerge without the foundational basis. Thus, the international regulatory efforts may not be congruent with each other. In turn, the variety of standards without shared semantic understanding may not necessarily serve as an effective regulatory approach to mitigate the risks associated with the development of AS. Thus, it is essential for the respective US departments to clarify their position towards such standards and accelerate the work where standardization is absent or in the nascent stage of development.

Table 3  Summary of human control over AS

|  | Type of control | Operational role | Intervention role | Key requirements |
|---|---|---|---|---|
| Direct control | Human-in-the-loop | Human | Human | (i)   The active presence of a qualified human operator; (ii)   system's responsiveness in completing tasks in a timeframe consistent with operator intentions |
|  | Human-on-the-loop | System | Human | (i)   The presence of a human operator; (ii)   a manual override feature that allows an operator to assume control of the AS at any time; (iii)   operator's relevant credentials |
|  | Remote Control | System/Human | Human | (i) Requirement to link systems with a qualified human operator; (ii) two-way communication links; (iii) monitoring information |
| Indirect control | Design control as default | System | System or human | (i)   Relevant software V&V, T&E; (ii)   compliance with relevant law; (iii)   malfunction notification technology; (iv)   wireless communications and system location technology; (v)   a data-recording system; (vi)   operator's relevant credentials |
|  | Control by "ethical code" | System | System | (i)   Same requirements as design control as default; (ii)   additional requirement in the form of ethical guidelines for IT developers |

This article postulates to steer public and regulatory debate away from the linguistic complexities of the term "autonomy" and focus instead on the areas where the human factor is undergoing important changes. This is where the standardization is required. Based on our research, it is important to differentiate between direct human control, which is concerned with certain degree of manual human intervention over the use of a system, and indirect control, which stipulates the design requirements of a system. The analysis of the transportation and defense sectors illustrates that both dimensions of control should be considered but one has to be mindful that each have different requirements. Standardization related to indirect control has to deal with procedures for measuring, testing, and certifying a machine's system abilities, particularly in the context of failure.[130] Furthermore, it has to consider the development of ethics certification programme for AI systems, specifically in relation to transparency, accountability and algorithmic bias.[131] Standardization related to direct control should have a different focus. It has to consider the requirements for human operators or supervisors of AI systems in various application domains, interoperability features, such as the specification of manual override function and associated best practices, as well as the standards of communication technologies. This area of analysis has received, so far, less consideration from main standard setting organizations that focus on AI. While it is understandable that regulators focus primarily on the software engineering processes of AI systems, our research shows that direct human control is still relevant even though its role is shifting more to human on-the-loop or remote control. In the context of AS, the "direct" element of human control is less conceived as a separate entity, but rather as an integral element of the socio-technical system. It means that direct human control works necessarily in conjunction with default control features. Thus, interoperability standard is particularly important. The potential standardization here can be divided into both technical and non-technical dimensions. Technical interoperability includes communications, electronic, application and database interoperability, while non-technical interoperability includes human-machine integration, as well as operational and process interoperability. This area of work requires further attention from regulatory bodies should one follow the dichotomy of control exemplified by the DoD and DoT experience with the requirement of human control.

## 6. Concluding remarks

In the current literature on AS there is an oft-repeated argument that humans should retain control over AS. Our research shows that the notion of human control is far from univocal. Various dimensions of human control and domain-specific requirements are necessary to ensure certain desirable levels of such control. There are at least two meaningful types of human control: (i) direct control related to the physical ability of a human to decide on the course action of a system; and (ii) indirect control which is stipulated at the level of design and relates to trust-building measures.

In defense, human control have two main dimensions: control in the form of a human finger on the button, and control-by-design. However, in some circumstances, the requirement of human control has been limited to control-by-design and defined in a constrained way that gives little confidence as to whether the requirement for an "appropriate level of human control" can actually be satisfied. In transportation, there are also two types of human control (manual control and control-by-design) with more explicit appreciation given to the ethical requirements at the level of design. However, the majority of states define requirements regarding manual control only. While the requirement for manual control is seen as a major obstacle towards the wider implementation of AVs, states such as California have adopted regulations which maintain elements of both direct and indirect control in the form of remote control and accompanying design requirements.

The typology of human control shed lights into the discussion on governance of AS more broadly by focusing on more detailed rules and standards applicable to the changing role of human factors. The standardization is considered as a particularly relevant regulatory tool to address ever-changing dynamics of technology development. Based on our findings, standardization related to indirect control has to focus on the one hand with measuring, testing, and certifying a machine's system abilities, particularly in the context of failure (control as a design default). On the other hand, it has to develop ethics certification programme, specifically in relation to transparency, accountability and algorithmic bias (control by ethical code).[132] Standardization related to direct control has to consider the requirements for human operators or supervisors of AI systems in various application domains, such as the specification of manual override function and associated best practices, as well as the standards of communication technologies. A particular issue that requires attention is the problem of non-technical interoperability, that is the integration of direct human control with default control features specified at the level of design.

## Acknowledgments

Araz Taeihagh is grateful for the support provided by the Lee Kuan Yew School of Public Policy, National University of Singapore.

## Endnotes

1. Bureau of the Investigative Journalism, "Drone Warfare" [Last accessed 28 August 2019.] Available from URL: http://www.thebureauinvestigates.com/category/projects/drones/.
2. Only recently, more attention has been paid to the ways in which artificial intelligence (AI) can be used maliciously and how it could affect various actors. See M Brundage et al. (2018).
3. See, among others, Anderson and Waxman (2013), Lewis (2015), and Schmitt (2013).
4. In legal theory, concepts such as "norms," "rules," or "principles" have various, often conflicting, meanings. While appreciating a rich legal scholarship, this article does not intend to unpack these concepts, as this task is not seen as necessary to the present study. The only caveat is that rules are considered in this study as more concrete specifications of a general norm of human control.
5. See, in the context of autonomous weapons, Ekelhof (2018) and Roorda (2015).
6. See Gordon (1980) and Dean (2010).
7. In this article, we use the words "policies" and "principles" as synonymous, although in the literature various authors emphasized the difference between these two concepts (see Dworkin 1977). Moreover, some organization such as International Committee of the Red Cross (ICRC) argues that "human control" should be even considered as a legal norm (see Chengeta 2018; ICRC 2018).
8. The problem of what level of human judgment is indeed "significant" and what terminology should be used to describe the level of human judgment is debated in the literature. Similarly, neither is there agreement on what the term "autonomy" actually means in the context of weapons, nor clear delineation between "autonomous," "semi-autonomous," "human-supervised," "automated," and "automatic" weapons.
9. DoD, Directive 3000.09 *Autonomy in Weapon Systems* (2012).
10. DoD, Directive 3000.09 *Autonomy in Weapon Systems* (2012).
11. DoT, *Preparing for the Future of Transportation: Automated Vehicles 3.0* (2018).
12. Nunes *et al.* (2019).
13. See Illinois State HB 791/2017, Connecticut SB 260/2017, or Washington D.C. DC B 19-0931/2012; see more on the National Conference of State Legislatures (NCSL) database. [Last accessed 28 August 2019.] Available from URL: https://www.ncsl.org/research/transportation/autonomous-vehicles-self-driving-vehicles-enacted-legislation.aspx.
14. Fischer (1982).
15. See the discussion on two models of autonomous targeting in Leveringhaus (2016).
16. Scharre and Horowitz (2015).
17. Scharre and Horowitz (2015).
18. Thomas (2015).
19. Marra and McNeil (2013).
20. Marra and McNeil (2013).
21. Taeihagh and Lim (2019).
22. Parasuraman *et al.* (2000).
23. Similarly see Air Force Research Lab 11-level autonomy ranking. See Sholes (2007).

24 An example of this is the Air Force Research Lab 11-level autonomy spectrum, which considers autonomy across various stages of machine action.
25 Roff (2015).
26 Roff (2015).
27 Howlett *et al.*, this issue, and Erdelyi and Goldsmith (2018).
28 DoD (n 9).
29 DoT (n 11).
30 UN, *Background on Lethal Autonomous Weapon Systems* [Last accessed on 28 August 2019.] Available from URL: https://www.un.org/disarmament/geneva/ccw/background-on-lethal-autonomous-weapons-systems.
31 DoT (n 11).
32 T Dye, *Top Down Policymaking* (CQ Press 2000).
33 Pülzl and Treib (2007).
34 Pülzl and Treib (2007).
35 Implementation of a policy relates to the actual effects of a given policy, which may only to certain extent reflect intended objectives. Here, a bottom-up or hybrid approach is more justified, as studies show that social outcomes do not always sufficiently relate to original policy plans (Pülzl & Treib 2007).
36 For the study of policy operationalization, see Kerwin and Furlong (2011).
37 Dye (2016).
38 See Lewallen, this issue, for an in-depth examination of challenges of delineating boundaries and spill-over effects of emerging technologies.
39 In the United States, the Administrative Procedure Act (1946) is the federal statute that governs the way in which administrative agencies of the federal government may propose and establish regulations.
40 DoD (n 9), The White House, National Security Strategy (2017, 2015, 2010), DoD, National Defense Strategy (2018); DoD, National Military Strategy (2015, 2011); DoD, Defense Strategic Guidelines (2012), DoD, The Quadrennial Defense Reviews (QDR 2014, 2010), DoD, Unmanned Systems Integrated Roadmap FY2013–2036 (2014) and FY2011–2036 (2011), DoT (n 11), DoT, NHTSA *Automated Driving Systems: A Vision for Safety 2.0* (Report) (2017), SAE, SAE J 3016-2018: *Taxonomy and Definitions for Terms Related to Driving Automation Systems for On-Road Motor Vehicles* (2018). NHTSA, *Federal Automated Vehicles Policy*, 2016.
41 DoT (n 11).
42 DoD Department of the Army, *Army Techniques Publication 3–60* (2015).
43 DoD JCIDS, *Manual for the Operation of the Joint Capabilities Integration and Development Systems* (2015).
44 Defense Science Board (DSB), *Summer Study on Autonomy* (Report) (2016).
45 NCSL (n 13).
46 ROE are directives to military forces that define the circumstances, conditions, degree, and the manner in which force, or actions, which might be construed as provocative, may be applied. See DoD CJCSI, 3121.01B *Standing Rules of Engagement/Standing Rules for the Use of Force for US Forces* (2005).
47 DoD, National Security Strategy (2017) 20; DoD, National Defense Strategy (2018) 3,7.
48 DoD (n 9) 1.a.
49 DoD (n 9) 1.a. PART II Definitions.
50 DoD (n 9) 1.a. 4.c. (1), 4.c. (3), 4.d.
51 DoD (n 9) 1.a. 4.a.
52 See Government of the United States, *Human-Machine Interaction in the Development, Deployment and Use of Emerging Technologies in the Area of Lethal Autonomous Weapons Systems* (Statement CCW/GGE.2/2018/WP.4) (2018). For further context see Article 36, *Killer Robots: UK Government Policy on Fully Autonomous Weapons* (2013) [Last accessed on 28 August 2019.] Available from URL: http://www.article36.org/wp-content/uploads/2013/04/Policy_Paper1.pdf.
53 Government of the United States (n 52).
54 DSB (n 44).
55 Roff and Danks (2018).
56 Roff and Danks (2018).

57 Roff and Danks (2018).

58 DoD, Unmanned Systems Integrated Roadmap FY2011–2036 (2011); see also Leveringhaus (2016).

59 DoD (n 9) 4.a.

60 DoD 1.a (2).

61 The United States already had a legal review mechanism in 1974 before Additional Protocol I to the Geneva Conventions came into force, which in Article 36 requires legal review of new weapons. See WH Parks, *Conventional Weapons and Weapons Reviews* (2005) Year Book of International Humanitarian Law.

62 See ICRC (2006) and Boothby (2016).

63 See M Schmitt (n 3); J Lewis (n 3).

64 See van den Boogaard (2015).

65 Farrant and Ford (2017).

66 DoD (n 58).

67 DoD, Unmanned Systems Integrated Roadmap FY2013–2036 (2014).

68 Joint warfare is a military doctrine, which places priority on the integration of the various service branches of a state's armed forces into one unified command.

69 DoD (n 42).

70 J Farrant and C Ford (n 65) 412–415.

71 DSB (n 44).

72 DoD (n 67).

73 DSB (n 44).

74 DoD (n 9) 2.

75 See N Sharkey, "Staying in the loop" in Nehal Bhuta *et al.* (eds) *Autonomous Weapon Systems* (CUP 2016) 23–38.

76 See N Sharkey (n 75); Cummings (2004).

77 Haslbeck and Hoermann (2016).

78 DoD (n 9) 2.

79 DSB (n 44).

80 DoT, NHTSA (n 40).

81 DoT (n 11).

82 DoT (n 11).

83 Collingwood (2017).

84 DoT (n 11).

85 DoT (n 11).

86 DoT (n 11).

87 SAE (n 40).

88 DoT (n 11).

89 N Boudette, *Tesla Self-Driving System Cleared in Deadly Crash*. [Last accessed 28 August 2019.] Available from URL: https://www.nytimes.com/2017/01/19/business/tesla-model-s-autopilot-fatal-crash.html.

90 J Davis, *Dreaming of Driverless: What's the Difference Between Level 2 and Level 5 Autonomy?* (2018). [Last accessed 28 August 2019.] Available from URL: https://blogs.nvidia.com/blog/2018/01/25/whats-difference-level-2-level-5-autonomy.

91 Cunningham and Regan (2017).

92 Narayanan (2019).

93 SA Beiker, *Legal aspects of autonomous driving* (2012) Santa Clara L. Rev. 1145.

94 Hübner and White (2018).

95 DoT (n 11) 49–63.

96 DoT, NHTSA (n 40).

97 DoT (n 11).

98 NCSL (n 13).

99 State of Illinois, Executive Order 2018–13 (2018).

been put forward, among others, by the UK National Endowment for Science, Technology and the Arts. See T. Snow, 'Three key principles to help public sector organisations make the most of AI tools' (Report) (2019).

132    IEEE SA recently launched the development of an Ethics Certification Program for Autonomous and Intelligent Systems (ECPAIS). See also ISO/IEC 24027: Bias in AI systems and AI aided decision making and ISO/IEC 24028: Overview of trustworthiness in Artificial Intelligence.